# Challenges concerning the discriminatory optical force for chiral molecules

David S. Bradshaw, Matt M. Coles and David L. Andrews[*]

*School of Chemistry, University of East Anglia, Norwich NR4 7TJ, United Kingdom*

## Abstract

In response to arXiv:1506.07423v1 we discuss the authors' work, and our own, on proposed schemes aiming to achieve a discriminatory optical force for chiral molecules.

## 1. Introduction

In connection with recent developments in the theory of chiral discrimination between molecules, we respond to a submission by Cameron, Barnett and Yao [1] that has aimed to defend those authors' previous works [2, 3], whilst also introducing a suggestion of error in some of our research on a different mechanism. The intention here is to scientifically argue each issue on a sound, mathematically consistent and logical basis, bringing clarity to the debate by focusing on the photonic mechanisms, openly responding to all of the comments raised in ref. [1]. One clarification to our description is made; none of our conclusions are changed by it, nor is any of our mathematical analysis.

It is appropriate to begin with inspection of the foundational equation (1) in ref. [1], which denotes a rotationally-averaged Lorentz force on a neutral, polarizable molecule:

$$\langle \overline{\mathbf{F}} \rangle = a \nabla w \pm |b| \nabla h + ... \qquad (1)$$

This expression is presented by the authors as the basis for describing an optical force, $\overline{\mathbf{F}}$, for chiral molecules. The material response parameter in the first term involves the trace of a molecular polarizability (entailing two electric dipole interactions, which we denote by E1$^2$) and the second, the trace of an analogous electric dipole-magnetic dipole (E1M1) tensor. We disagree that the latter represents 'mutual interference of electric-dipole / magnetic-dipole transition moments', because it does not involve interference but a sum of products. It is noteworthy that, in ref. [1], there is no mention of electric quadrupoles in connection with equation (1) despite a statement, later in the same paper, that the 'neglect of electric-quadrupole contributions cannot be justified'; a related paper from another group [4] also excludes the electric quadrupole contribution. The authors do, however, make an assertion that both magnetic–dipole and electric–quadrupole couplings contribute to the $|b|\nabla h$ term in an earlier work [3]. To clarify the issues arising from the manner of application of equation (1) in ref. [1], and our own understanding, it will be helpful to go straight to the heart of the process and consider in detail the role of the quantized radiation field.

## 2. Photon-based perspective

There are two forms of optical mechanism that can result in the centre of mass motion of a free, uncharged, polarizable particle. In one, relating to the dynamic Stark shift, the particle's interaction with light produces intensity-dependent internal energy level shifts, introducing an effective potential energy surface wherever the light intensity varies with the particle position (most familiarly across the dimensions of a beam, as in optical tweezers). In this case the radiation field suffers no change in state, and the position-dependent particle energy can be evaluated as the expectation value of an interaction term which delivers the expression $\alpha E^2$ (where $\alpha$ is the polarizability tensor and **E** the optical electric field). At a quantum level, this interaction involves photon annihilation and recreation into the same radiation mode – which can be cast as forward Rayleigh scattering in a Feynman diagrammatic sense. This is the origin of our scheme [5-9], and the mechanism is also the basis for numerous other works on optical forces. For example, it is consistent with optical lattice experiments in which the variation of internal energy with position serves as a potential energy surface and produces an optical dipole force. We shall refer to interactions of this kind as the first mechanism.

---

[*] Email: david.andrews@physics.org

A second mechanism describing centre of mass motion comprises processes within which the radiation state suffers change, through a beam deflection and/or frequency shift, that is quantum mechanically (*i.e.* always in principle) observable. At the photon level this signifies the occurrence of scattering events in which there is an annihilation of photons in one mode, and the creation of photons in another mode with a different wave-vector. Accordingly the radiation field as a whole suffers a change in momentum; the corresponding momentum difference is imparted to the particle. The rate at which this process occurs determines the rate of change of momentum and hence the force, as in the familiar formula for radiation pressure. This is the basis for early research done by Ashkin [10] and laser cooling techniques including optical molasses [11]. The difference in outcome, for the radiation, is such that no individual photon from a given optical input can manifest both types of mechanism, since the final radiation state cannot be both the same and different from the initial state.

Although it will be shown that our conclusions remain unchanged, we need to rectify one statement in the conference paper [12], quoted in full on p.3 in ref. [1]. This was an oversimplification of the issues, intended to deliver an audience the most readily comprehensible explanation; it should have been made clear that it only applies to the first mechanism described above. A more concise statement appears in our journal article [5] – in which we state 'since the initial and final states are typically not identical, energies and forces ought not to arise from necessarily off-diagonal matrix elements' – which is true insofar as it relates to the evaluation of dynamical energy shifts whose evaluation requires expectation values for the radiation field. If a process with identifiably different radiation modes is involved – as, for example, in Doppler cooling – then the ensuing changes in particle motion are the result of a process that occurs at a well-defined rate. This involves the conversion of photons from one radiation mode into another. The crux is this; when the initial and final states differ in this way, the process is incoherent and, therefore, a rotational average has to be performed on the *square* modulus of the quantum amplitude [13, 14].

Before returning to the former mechanism, it is helpful to dispense with any prospect of the second kind generating chiral discrimination. To exhibit such behaviour, the second mechanism would have to invoke an interference of amplitudes associated with **α** (E1$^2$) contributions with those involving the **G** tensor (E1M1), and also with those involving the **A** tensor (E1E2): the tensors are explicitly defined in ref. [5]. To account for the response of molecules in any fluid phase requires the evaluation of three-dimensional tensor rotational averages, the procedures for which are well established. Accordingly the interference terms would here introduce averages of fourth and fifth rank, respectively, which take the form of linear combinations of isotropic tensors of equivalent rank [15-17]. The corresponding physical observable could not be represented by the form depicted by equation (1), in which the trace of the molecular polarizability arises from a second-rank rotational average; this emphasizes that the scheme suggested in refs [2, 3] is not related to the incoherent, radiation field-changing mechanism. In fact, even when the relevant rotational averages are applied to a two-beam system with orthogonal linear polarizations, it becomes clear that any term sensitive to molecular chirality will vanish. This is because each product of an appropriate rank isotropic tensor with a salient member of the vector set {$\mathbf{e}_1$, $\mathbf{b}_1$, $\mathbf{k}_1$, $\mathbf{e}_2$, $\mathbf{b}_2$, $\mathbf{k}_2$} gives zero. (Here, $\mathbf{e}_{1/2}$ and $\mathbf{b}_{1/2}$ are the electric and magnetic polarizations of beam 1 or 2; $\mathbf{k}_{1/2}$ denotes the corresponding wave-vector.) We conclude that it could only be the first mechanism, detailed above, that represents a conceivable basis for the envisaged two-beam chiral separation although, unfortunately, we also determine a zero result for the system proposed [2,3]; this is explained below.

Now we return to the first mechanism. As observed above, the scalar molecular polarizability $\alpha$, relating to the first term of equation (1), represents the trace of a tensor form which involves a product of transition electric dipole moments, which on quadratic coupling with the electric field, $\mathbf{E}$, of the radiation gives as numerator the square of an energy term, and an energy denominator – so that an energy shift is delivered by the expression $\alpha\mathbf{E}^2$. Similarly the electric dipole-magnetic dipole (E1M1) tensor, or **G** tensor, corresponds to the second term of equation (1); we shall deal subsequently with any contribution from an electric quadrupole. The interaction Hamiltonian, $\hat{H}_{int}$, is an energy operator that, when assigned to an E1 interaction and cast between different molecular states, delivers a result containing a real electric dipole transition moment (given real wavefunctions, which can always be chosen). When the off-diagonal matrix elements of $\hat{H}_{int}$ are evaluated for M1 interactions, for coupling with the magnetic field, $\mathbf{B}$, of the radiation, a purely imaginary magnetic dipole transition moment is generated. Hence, whilst $\alpha\mathbf{E}^2$ is quite correctly a real energy, **GEB** gives an imaginary quantity if the two electromagnetic fields are linearly polarized, *i.e.* when their polarization vectors are expressible as real quantities.

We suggest that, for the interpretation of equation (2) of ref. [1] as a physically observable energy shift, it must deliver a real result;

$$\Delta W = \sum_{j \neq i} \frac{\langle i|\hat{H}_{int}|j\rangle\langle j|\hat{H}_{int}|i\rangle}{W_i - W_j}. \qquad (2)$$

Here $i$ is the identical initial and final state, $j$ is an intermediate state, $\Delta W$ denotes the energy shift (effective potential energy) and $W$ is the energy of a system state denoted by its subscript. It is readily established that the **GEB** contribution is null when linearly polarized light is used – since, as stated earlier, the response tensor is purely imaginary and the polarization vectors are real. The energy associated with this term, and hence also any derivative force, is imaginary. In contrast, our analysis [5] deploys circularly polarized light which relates **E** and **B** to complex polarization vectors of the form $\mathbf{e}^{(L|R)} = (1/\sqrt{2})(\hat{\mathbf{i}} \pm i\hat{\mathbf{j}})$ and $\mathbf{b}^{(L|R)} = (1/\sqrt{2})(\hat{\mathbf{j}} \mp i\hat{\mathbf{i}})$ (where $\hat{\mathbf{i}}$ and $\hat{\mathbf{j}}$ are mutually orthogonal unit vectors perpendicular to the direction of light propagation, and the superscripts denote left- or right-handed polarized light), respectively, so that the **GEB** contribution is non-zero; this correlates with the fact that this form of polarization is more often employed in chirality studies.

Using similar arguments, we conclude that the electric dipole-electric quadrupole contribution can equally be omitted from consideration. This is fully explained in our previous work [5]; it is again connected to the notion that the energy shift is a physical observable and must correspond to a real quantity.

At this juncture we note an objection, by the authors of ref. [1], to the omission of the electric quadrupole contribution. Of course, both the electric-quadrupole and magnetic-dipole forms of interaction emerge from the multipolar canonical transformation via the $\nabla \cdot \boldsymbol{a}$ term, where $\boldsymbol{a}$ is a vector potential [18]. Moreover, in all the symmetry point groups that support molecular chirality (the pure rotation groups) all transitions that are both E1 and M1 allowed are also E1 and E2 allowed – and vice-versa. So in establishing the principle of a non-vanishing chiroptical phenomenon, it suffices to show that one contribution is non–zero. The only exception to this logic would be if the M1 and E2 transition contributions delivered effects of equal and opposite magnitude. However, this condition could only arise in highly exceptional cases, since there is no systematic proportionality between the electric-quadrupole and magnetic-dipole transition moments. Therefore, the rigor of our proposed scheme [5] does not suffer if the electric quadruple contribution is non-zero: in fact the optical force would likely be greater than we have specified. The secondary issue of invariance to choice of coordinate origin is not relevant; any dependence on the position of origin in a 3D-rotating molecule has meaning only in terms of the internal coordinates, not the spatial position of the molecule as a whole with respect to the laboratory frame – and only the latter is of concern in optical forces.

## 3. Response to other issues

Without providing justification, it is asserted in Section 2 of ref. [1] that our research is incompatible with the Lorentz force law, but it is in fact fully compatible; moreover, the assertion is inconsistent with the claim that our results lead to the optical force predicted by their system, based on the Lorentz force. Our own scheme does not require two beams, and the requirements for standing waves and beam inteference do not apply to our work; equations (3)-(6) of ref. [1] have no bearing on our results. The claim that our result has already been reported in a correct form, cast as equation (8), cites a thesis published after our original paper [5]. The authors of ref. [1] state that their system forms a basis for chiral diffraction gratings: we cannot (and did not) make any such claim for our single-beam arrangement – in short, the two proposed schemes are separate and clearly distinguishable from each other.

The aim of the numerical section of our papers is to determine an approximate value for the optical force. Equation (9) of ref. [1] represents this method of estimation: the division of the trace of **α** (which is characterised as $4\pi\varepsilon_0 d^3$) by the trace of **G** (divided by $c$, so that the units match) gives roughly the inverse of the fine structure constant. The value of the molecular dimension, $d$, used in our calculations is 10 nm; the authors of ref. [1] claim this to be an 'enormous' quantity. This value certainly signifies a large molecule, but there are numerous classes of molecule that can be of such a size – dendrimers afford an illustration. For example, the 8$^{th}$ generation polyamideamine dendrimer [19] has a diameter of 9.7 nm; it has ~100,000 degrees of freedom and it is transparent in the visible range. Suitable outer surface functionalization will certainly produce a chiral molecule with parameter values similar to those we chose. Other possible species on this scale, also with interesting photophysics, are chiral nanospheres such as those in which a gold core is covered by a sugar shell [20]. If we purposely choose a small molecule with a diameter of 1 nm, the optical force that we calculate is diminished by three orders of magnitude. However, at no point have we claimed that such a chiral force will be anything but very small, in fact we wrote previously [5]: 'The method we have described, as well as the others cited by way of comparison, clearly suffer severe limitations on throughput, likely to render commercial application economically impractical'.

## 4. Conclusion

All of our own calculations and the conclusions in refs [5-9] are correct. They are not, in any way, changed if superposition states of the kind suggested in equation (3) of ref. [1] are entertained. The claim that we 'barely acknowledged' the authors of ref. [1] is mistaken – we cited them in every work on the subject – and critically in only one journal article [5] and one proceedings paper [12]. We were also careful to recognise a range of other related papers in our publications [4, 21-42]; our scheme was indeed inspired by other recent research, including the articles by Cameron *et al.* [2] and also by Canaguier-Durand *et al.* [4]. Nonetheless, the phenomenon described in our own work is novel and operates on an entirely different basis from any two-beam system. Our analysis has shown that, in this field especially, it is crucially important to account for key features of the fundamental photon-molecule interactions, correctly handling issues of group theoretical symmetry and tensor construction. We fully endorse the sentiment that mistakes should not continue to propagate.

**Acknowledgments**

We are grateful to the Leverhulme Trust for funding our research. We also thank Dr Jamie Leeder and Mathew Williams for helpful comments.